\documentclass[journal]{IEEEtran}

\ifCLASSINFOpdf
\usepackage{listing}
  \usepackage[pdftex]{graphicx}
	\usepackage[pdftex]{hyperref}
	\usepackage{booktabs}
	\usepackage{amsmath}
	\usepackage{slashbox}
		\usepackage{textcomp}
	\usepackage{multirow}
	\usepackage[euler]{textgreek}
\usepackage{siunitx}
\newcolumntype{P}[1]{>{\centering\arraybackslash}p{#1}}
\newcolumntype{M}[1]{>{\centering\arraybackslash}m{#1}}

\usepackage[hang,font=small]{caption}%
	\usepackage{color,soul}
	
	\definecolor{gray}{rgb}{1,0.6,0.5}
   \usepackage[square, comma, sort&compress,numbers]{natbib}
   \graphicspath{{./figures/}}
\else
\fi
\allowdisplaybreaks

\begin{document}
\title{Video Compression with CNN-based Post Processing}

\author{Fan~Zhang,~\IEEEmembership{Member,~IEEE},
				Di Ma,	Chen Feng,
        and~David~R.~Bull,~\IEEEmembership{Fellow,~IEEE}
\thanks{The authors would like to acknowledge funding from UK EPSRC (EP/L016656/1 and EP/M000885/1) and the NVIDIA GPU Seeding Grants. We also appreciate the valuable advice provided by Dr. Debargha Mukherjee on AV1 coding configuration.}
\thanks{Fan Zhang, Di Ma,	Chen Feng, and David R. Bull are with the Bristol Vision Institute, University of Bristol, Bristol, UK. }}
\maketitle

\begin{abstract}
In recent years, video compression techniques have been significantly challenged by the rapidly increased demands associated with high quality and immersive video content. Among various compression tools, post-processing can be applied on reconstructed video content to mitigate visible compression artefacts and to enhance overall perceptual quality. Inspired by advances in deep learning, we propose a new CNN-based post-processing approach, which has been integrated with two state-of-the-art coding standards, VVC and AV1. The results show consistent coding gains on all tested sequences at various spatial resolutions, with average bit rate savings of 4.0\% and 5.8\% against original VVC and AV1 respectively (based on the assessment of PSNR). This network has also been trained with perceptually inspired loss functions, which have further improved reconstruction quality based on perceptual quality assessment (VMAF), with average coding gains of 13.9\% over VVC and 10.5\% against AV1. 
\end{abstract}

\begin{IEEEkeywords}

CNN, VVC, AV1, video compression, perceptual quality, GAN
\end{IEEEkeywords}

\IEEEpeerreviewmaketitle

\section{Introduction}
\label{sec:intro}

{The importance} of video compression has come to the fore over the past decade driven by the tension between the huge quantities of video content consumed everyday and the bandwidth available to transmit it. This challenge has been addressed through the creation of new video coding standards, the latest activity being by the Joint Video Exploration Team (JVET), who published the first version of H.266/Versatile Video Coding (VVC) \cite{s:VVC} in 2020. Compared to its predecessor, H.265/High Efficiency Video Coding (HEVC), VVC has achieved up to 50\% performance improvement through the adoption of numerous sophisticated coding tools, in particular with improved support for formats with high spatial resolutions, high dynamic range and spherical content. Alongside VVC, the Alliance for Open Media (AOMedia) also published its first video coding format, AOMedia Video (AV1) in 2018, which has been reported to offer comparable coding performance to VVC \cite{j:Chen1}.   

Machine learning, especially based on deep convolutional neural networks (CNNs), has been increasingly applied in the context of video compression and has achieved promising results both when used in conjunction with  conventional coding algorithms and in the form of new end-to-end architectures. In addition to conventional normative coding tools, deep learning can also be employed at the video decoder as a post-processing stage to further reduce noticeable artefacts and enhance the visual quality of compressed content. For the state-of-the-art coding standards, such as VVC and AV1, most existing learning-based post-processing approaches can only offer evident improvement for less efficient coding configurations (e.g. intra coding), and the employed networks are normally trained to minimise average absolute pixel distortions rather than explicitly to improve perceptual quality. 

Based on the Generative Adversarial Network (GAN) paradigm, we propose a novel CNN-based post-processing approach with an extension that achieves improved perceptual reconstruction quality. This approach has been evaluated on the VVC Test Model (VTM) 4.0.1 and on AV1 libaom 1.0.0, with results showing consistent improvement on standard JVET test sequences for different QP values based on different quality measurements. We have further analysed the computational complexities of different CNN structure variants and correlated them with overall coding gains.

This paper is a comprehensive extension of our previous work \cite{c:Zhang30}, which solely focused on the PSNR driven optimization of VVC compressed content. The primary differences are summarized below:

\begin{itemize}
	\item The CNN model used for post-processing has been extensively upgraded with a new GAN-based perceptual training strategy, which can significantly improve the perceptual quality of the final reconstructed content compared to \cite{c:Zhang30}.
	\item This CNN-based post processing approach has been also trained and evaluated on AV1 compressed results (alongside VVC in \cite{c:Zhang30}) and achieved similar coding gains. 
\end{itemize}

In the remainder of the paper, we first survey the prior work in the field of deep video compression and, in particular, describe learning-based post-processing approaches. Secondly, we present our proposed CNN-based post-processing approach, describing the network architecture and how it was trained and evaluated. We then summarize and discuss the experimental results, highlighting the performance improvements over standard video codecs. Finally we conclude the paper and outline possible future work.

\section{PRIOR WORK}
\label{sec:review}

\subsection{Deep video compression}

In the past few years, machine learning, in particular deep neural networks, has been increasingly applied to image and video compression, demonstrating significant potential compared to conventional coding methods. Learning-based video coding algorithms can be classified into two primary groups: new end-to-end network architectures and those that enhance individual conventional coding tools.

Machine learning has been employed to refine existing coding tools within a conventional coding framework, including intra and inter prediction, transformation, quantization and in-loop filtering \cite{j:Liu3}. New coding tools have also been developed with the support of neural networks, such as CNN-based spatial resolution and bit depth adaptation \cite{j:Zhang12}. Moreover, the classic hybrid video coding framework has been challenged by new deep network architectures, which enable end-to-end training and optimization \cite{c:Balle,c:Rippel}. This latter approach often employs a general rate distortion framework with non-linear transforms, which are based on convolutional filters and nonlinear activation functions. Although these solutions show great promise as an alternative to conventional codecs, their performance cannot still compete with the latest standardized video codecs, including VVC and AV1.

\subsection{CNN-based post processing}

Compression processes often introduce various visible artefacts such as blocking mismatches, banding and blurring, especially when large quantization steps are employed. These unpleasant distortions can be mitigated by filtering the reconstructed frames. When this enhancement process is performed outside of the encoding loop (generally after decoding), it is referred to as  post-processing. 

In standardized codecs, filters have also been designed for use within the encoding loop to reduce compression artefacts. VVC employs three different types in-loop filters including de-blocking filters (DBF), sample adaptive offset (SAO) and adaptive loop filters (ALF) \cite{s:VVC}. In AV1, in addition to deblocking filters, there are two other filtering operations:  constrained directional enhancement filtering and loop restoration filtering \cite{j:Chen1}.

CNNs are now also playing an important role in image restoration (including super-resolution), and these approaches can also be employed for post-processing of compressed video content to improve the overall reconstruction quality. Although various researchers have implemented CNN-based post-processing approaches in the context of HEVC and VVC \cite{j:Ma2,j:Zhao3,r:JVETN0254,r:JVETO0079}, most of these can only achieve coding gains for All Intra configurations, which offer lower coding efficiency compared to Random Access configurations based on hierarchical B frame structures. In addition, all the employed CNN models in these approaches were trained to optimize a simple loss function based on pixel distortions ($\ell 1$ or $\ell 2$ loss), which can lead to over-smoothed reconstruction results. 

\section{PROPOSED ALGORITHM}
\label{sec:algo}

\begin{figure}[htbp]
\centerline{\includegraphics[width=1.01\linewidth]{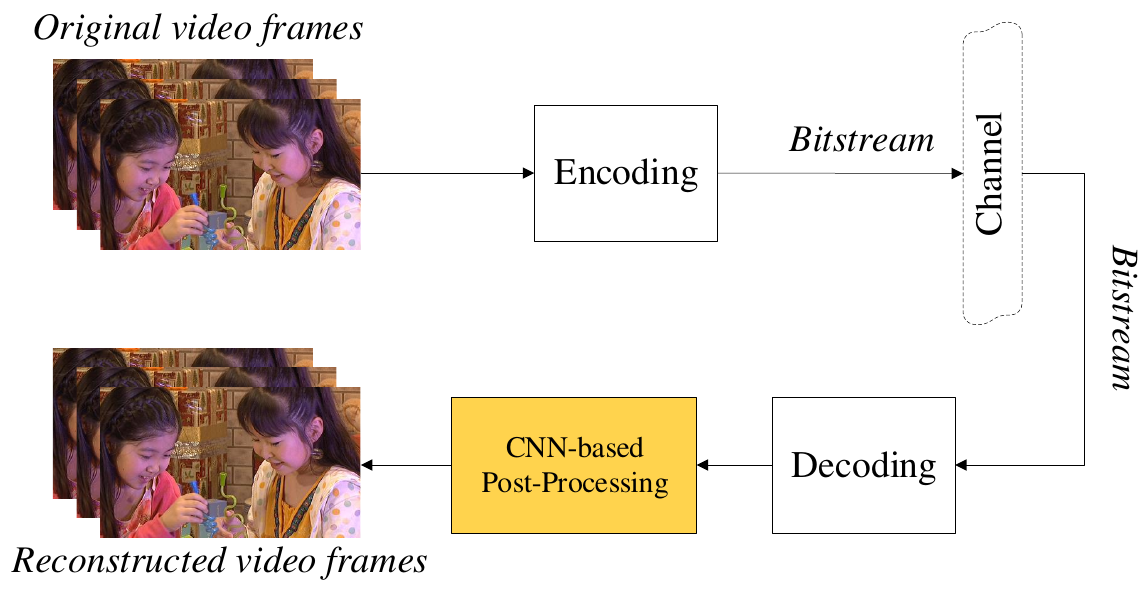}}
\caption{A typical coding workflow with a CNN-based post-processing module.}
\label{fig:workflow}
\end{figure}

{\bf Figure \ref{fig:workflow}} shows a high level coding workflow with a CNN-based post-processing module.  This section focuses on the structure of the employed CNN architecture, and the details of network training and evaluation.

\subsection{The CNN architecture}

\begin{figure*}[htbp]
\centerline{\includegraphics[width=1.01\linewidth]{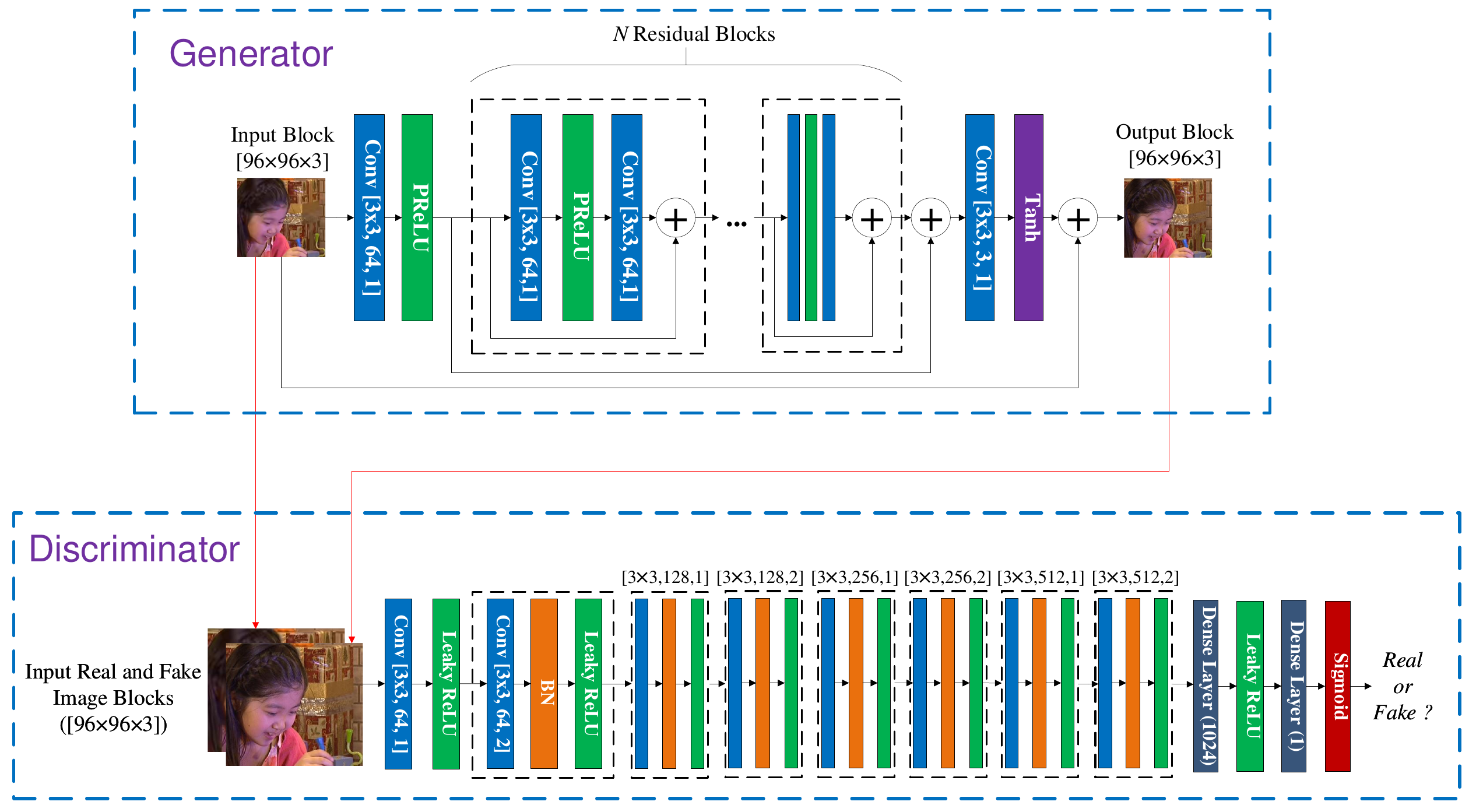}}
\caption{The employed GAN architecture, comprising generator and discriminator stages}
\label{fig:gan}
\end{figure*}

The CNN architecture used in this work is illustrated in {\bf Figure \ref{fig:gan}}. 

\subsubsection{The generator network} is a modified version based on the generator (SRResNet) of SRGAN \cite{c:Ledig}, and this has been previously employed by the authors in video compression systems based on  spatial resolution and bit depth resampling \cite{j:Zhang12}. This network takes a 96$\times$96 RGB (compressed) image block as input and produces an image block with the same format, targeting to its corresponding original (uncompressed) counterpart\footnote{We have used the same input/output formats as in SRResNet.}. 

Residual blocks (RB) form the basic unit in this network, which contains two convolutional layers and a parametric Rectified Linear Unit (PReLU) activation function in between them. A skip connection is used between the input of each RB and the output of its second convolutional layer. The number of residual blocks is configurable and was set to 16 in this work. 

The input of the network is connected to these successive RBs through a convolutional layer (also with a ReLU). Between the network output and the output of the last RB, there is also a convolution layer (output layer) followed by a Tanh activation function. An additional skip connection is used between the output of the input layer and the output of the last RB. A long skip connection is also employed between the input the first RB and the output of the output layer to produce the final output. 

\subsubsection{The discriminator} used in this work is similar to that in SRGAN \cite{c:Ledig}, which takes the output of the generator (\textit{fake}) and compares to its corresponding original (\textit{real}). This network consists of one input layer (with a Leaky ReLU), seven identical convolutional layers, and two dense layers. Each of the convolutional layers is followed by a batch normalization layer and a leaky ReLU activation function. After the second dense layer, a Sigmoid activation function is employed to output a probability to predict how much the quality of the \textit{real} image block is perceptually better than the \textit{fake} one.

The parameters used in each convolutional layer including kernel sizes, feature map numbers and stride values are shown in Figure \ref{fig:gan}.

\subsection{Training database}

The training material is essential for learning-based algorithms. We need to ensure the training content is diverse and covers various texture types in order to achieve good model generalisation and avoid potential over-fitting problems. To train the employed network, we have selected 432 uncompressed video sequences from a publicly available training database, BVI-DVC \cite{j:Zhang17}, which was designed specifically for deep video compression. All these sequences have the same frame rate of 60 frames per second, YCbCr 4:2:0 format, and with four different spatial resolutions including 3840$\times$2160, 1920$\times$1080, 960$\times$540, and 480$\times$270. We have encoded these 432 original sequences using VVC VTM 4.0.1 and AV1 libaom 1.0.0 with the coding configurations summarized in \textbf{Table \ref{tab:cfg}}.

\begin{table*}[ht]
 \caption{The coding configuration employed for VVC VTM and AV1 libaom.}
\centering 
		\begin{tabular}{c | p{1.6cm} | p{10.7cm}}
		\toprule
	{Codec}	& {Version} & {Configuration parameters} \\
				\midrule
VVC VTM & 4.0.1 & Random Access configuration \cite{s:JVETCTC}. IntraPeriod=64, GOPSize=16, QP=22, 27, 32, 37, 42.\\
				\midrule
AV1 libaom& 1.0.0-5ec3e8c  (02/05/2020)&  {-{}-}i420 {-{}-}psnr {-{}-}usage=0 {-{}-}verbose {-{}-}cpu-used=0 {-{}-}threads=0 {-{}-}profile=0 {-{}-}width=\$w  {-{}-}height=\$h {-{}-}input-bit-depth=10 {-{}-}bit-depth=10 {-{}-}fps=\$fps/1001 {{-{}-}passes=1} {{-{}-}kf-max-dist=64 {-{}-}kf-min-dist=64 }{-{}-}drop-frame=0 {-{}-}static-thresh=0 {-{}-}arnr-maxframes=7 {-{}-}arnr-strength=5 {{-{}-}lag-in-frames=19} {-{}-}aq-mode=0 {-{}-}bias-pct=100 {-{}-}minsection-pct=1 {-{}-}maxsection-pct=10000 {-{}-}auto-alt-ref=1 {-{}-}min-q=0 {-{}-}max-q=63 {{-{}-}max-gf-interval=16} {-{}-}min-gf-interval=4 {-{}-}frame-parallel=0 {-{}-}color-primaries=bt709 {-{}-}end-usage=q {-{}-}sharpness=0 {-{}-}undershoot-pct=100 {-{}-}overshoot-pct=100 {-{}-}tile-columns=0 {-{}-}cq-level=\{32, 43, 55, 63\} {w/o} {-{}-}enable-fwd-kf=1\\
\bottomrule
\end{tabular} 
\label{tab:cfg}
\end{table*} 

For each codec, the reconstructed video frames for each QP value and their corresponding originals were randomly selected and segmented into 96$\times$96 colour image blocks (after converting to the RGB space from YCbCr 4:2:0). We have also rotated selected image blocks to further improve data diversity. As a result, for each video codec and QP group\footnote{Based on our preliminary results, we note that, through QP sub-grouping, additional coding gains (approximately 0.05dB based on PSNR) can be achieved compared to using a single CNN model.}, there are over 100,000 image blocks pairs (compressed and original).

\subsection{Training strategy}

We trained this network using two different methodologies: (i) only train the generator using  $\ell 1$ loss (mean absolute difference)  (ii) jointly train both the generator and the discriminator based on perceptually-inspired loss functions. The CNN models obtained by these two training methods are used to post-process VVC and AV1 compressed content in the evaluation experiments, and their results are compared in the next section.

\subsubsection{$\ell 1$ loss}

is firstly employed to train the network (generator only) using the material generated for each QP group and codec. This results in a total number of nine CNN models for different evaluation scenarios:

\small
\begin{equation}
\left\{ 
\begin{array}{l r}
\mathrm{CNN}_\mathrm{VVC,QP22}, \  \text{if} & \mathrm{QP_{eval}}\leq 24.5 \\
\mathrm{CNN}_\mathrm{VVC,QP27}, \  \text{if} & 24.5<\mathrm{QP_{eval}}\leq 29.5 \\
\mathrm{CNN}_\mathrm{VVC,QP32}, \  \text{if} & 29.5<\mathrm{QP_{eval}}\leq 34.5 \\
\mathrm{CNN}_\mathrm{VVC,QP37}, \  \text{if}  & 34.5<\mathrm{QP_{eval}}\leq 39.5 \\
\mathrm{CNN}_\mathrm{VVC,QP42}, \  \text{if}  & \mathrm{QP_{eval}}> 39.5
\end{array}
\right.
\end{equation}
\begin{equation}
\left\{ 
\begin{array}{l r}
\mathrm{CNN}_\mathrm{AV1,QP32}, \  \text{if} & \mathrm{QP_{eval}}\leq 37.5 \\
\mathrm{CNN}_\mathrm{AV1,QP43}, \  \text{if} & 37.5<\mathrm{QP_{eval}}\leq 49 \\
\mathrm{CNN}_\mathrm{AV1,QP55}, \  \text{if} & 49<\mathrm{QP_{eval}}\leq 59 \\
\mathrm{CNN}_\mathrm{AV1,QP63}, \  \text{if}  & \mathrm{QP_{eval}}> 59
\end{array}
\right.
\end{equation}
\normalsize

Here $\mathrm{QP_{eval}}$ represents the base QP value employed in the evaluation phase for the two different codecs, and $\mathrm{CNN}_{c,q}$ is the CNN model trained for different codecs (VVC or AV1) and QP values. 

\subsubsection{Perceptual loss functions}

have been employed to train the whole GAN architecture following a two stage training strategy. This was initially designed to train the Relativistic GAN for  image generation \cite{j:Jolicoeur}, and has also been used to train the CNN models for spatial resolution and bit depth up-sampling \cite{c:ZHang31,c:Zhang29}. In the first stage, the generator is trained separately using the multi-scale structural similarity index (MS-SSIM) \cite{c:mssim} as the loss function. The trained generator model is employed as the starting point when  the generator and discriminator are trained together in the second phase. 

The generator is trained using a combined loss function, $\mathcal L_\mathrm{gen}$ in the second stage:
\begin{equation}
      \mathcal L_\mathrm{gen} =\mathcal L_{\rm SSIM}+\alpha \cdot \mathcal L_{\ell1} + \beta \cdot \mathcal L_G^{a}
\label{eq:gen}
\end{equation}
in which $\mathcal L_{\rm SSIM}$ stands for the SSIM \cite{j:ssim} loss (1-SSIM) between the generator output and the target, while $\mathcal L_{\ell1}$ is the $\ell 1$ loss between them. $\mathcal L_G^{a}$ is defined as the adversarial loss for the generator:

\begin{small}
\begin{equation}
\begin{split}
      {\mathcal L_G^{a}}=&-E_{I_r}[{\rm ln}(1-({\rm Sig}(O_d(I_r)-E_{I_f}[O_d(I_f)])))]\\
			&-E_{I_f}[{\rm ln}({\rm Sig}(O_d(I_f)-E_{I_r}[O_d(I_r)]))]
\label{eq:alg}
\end{split}
\end{equation}
\end{small}

Here $I_r$ and $I_f$ are denoted as the \textit{real} and \textit{fake} image blocks respectively. $E_{I_r}[\cdot]$ represents the mean operation for all the \textit{real} (\textit{fake} if $I_r$ is replaced by $I_f$) image blocks, and $O_d(\cdot)$ is the output of the discriminator. `Sig' represents the Sigmoid function.

For the discriminator, the loss function $\mathcal L_D$ is given by (\ref{eq:1}):
\begin{small}
\begin{equation}
\begin{split}
      {\mathcal L_D}=&-E_{I_r}[{\rm ln}({\rm Sig}(O_d(I_r)-E_{I_f}[O_d(I_f)]))]\\
			& -E_{I_f}[{\rm ln}(1-({\rm Sig}(O_d(I_f)-E_{I_r}[O_d(I_r)])))]
\label{eq:1}
\end{split}
\end{equation}
\end{small}
\subsubsection{The training configuration} is summarized as follows. We implemented all networks based on the TensorFlow 1.8.0 framework, and set the learning rate and weight decay to 0.0001  and 0.1 (for every 100 epochs) respectively for both training stages. The total number of training epochs are 200. We have used Adam optimisation algorithm during the training with the hyper parameters of $\beta_1=0.9$ and $\beta_2=0.999$. The two weights $\alpha$ and $\beta$ in equation (\ref{eq:gen}) are set up to 0.025 and \num{5e-3} respectively.

\subsection{Evaluation operation}

In the evaluation stage, when we use the trained CNN models to enhance compressed video frames, each frame is segmented into 96$\times$96 overlapping image blocks with an overlap of 4 pixels. These blocks are converted to RGB color space as the input of CNN models, and the output image blocks are then aggregated in the same way through simple blending to form the final video frame. Here only the generator network is used in the evaluation (the discriminator is for training only).  

\section{RESULTS AND DISCUSSION}
\label{sec:results}

\begin{table*}[htbp]
\centering
\small
\caption{The compression performance of the proposed method benchmarked on the original VVC VTM 4.0.1.  Negative BD-rate values indicate coding gains.}
\begin{tabular}{l ||r |r ||r |r ||r |r ||r |r  }
\toprule
Method	& 	\multicolumn{4}{c||}{VTM-PP ($\ell 1$ loss)} &\multicolumn{4}{c}{VTM-PP (Perceptual loss)}\\
\midrule Metric&\multicolumn{2}{c||}{PSNR} &\multicolumn{2}{c||}{VMAF} &\multicolumn{2}{c||}{PSNR} &\multicolumn{2}{c}{VMAF}\\ 
\midrule QP Range	  &   {H-QPs} 		&    {L-QPs}	&   {H-QPs} &    {L-QPs}&{H-QPs} &{L-QPs} &{H-QPs} &{L-QPs}\\
\midrule Class-Sequence & BD-rate & BD-rate & BD-rate & BD-rate & BD-rate& BD-rate&BD-rate&BD-rate\\
\midrule
A1-Campfire&-3.3\% & -2.3\% &-5.6\% & -4.6\%  &+0.2\%&-0.4\%&-10.4\%& -10.7\%\\ 
A1-FoodMarket4&-2.6\%  & -2.0\%  &-3.8\%  & -3.0\% &-0.0\%&+0.1\%&-8.4\%& -7.4\%\\ 
A1-Tango2&-3.3\%  & -2.9\%  &-3.4\% & -3.0\% &-1.1\%&-0.6\%&-7.8\%& -9.3\%\\ 
\midrule A2-CatRobot1&-5.2\%  & -5.2\%  &-4.6\%  & -4.4\% &-0.6\%&-1.1\%&-14.4\%&-17.8\% \\ 
A2-DaylightRoad2&-6.0\%  & -7.1\%  &-6.8\% & -7.2\% &-1.1\%&-2.1\%&-19.5\%& -23.4\%\\ 
A2-ParkRunning3&-0.8\% & -0.4\%  &-2.3\% & -0.2\% &+2.1\%&+2.4\%&-11.7\%& -11.1\%\\ 
\midrule \textbf{Class A}&-3.5\% & -3.3\% &-4.4\%& -3.7\% &-0.1\%&+0.3\%&-10.1\%&-13.3\%\\ 
 \midrule B-BQTerrace&-2.2\%  & -1.0\% &-6.1\% & -1.1\% &+2.0\%&+0.3\%&-23.3\%& -28.8\%\\ 
B-BasketballDrive&-3.4\%  & -3.1\%  &-1.8\%  & +2.7\%&-0.9\%&-0.9\%&-7.9\%& -8.7\%\\ 
B-Cactus&-3.4\%  & -3.0\%  &-5.1\%  & -4.4\% &+0.2\%&-0.2\%&-15.8\%&-17.1\%\\ 
B-MarketPlace&-2.6\%  & -2.3\%  &-4.8\% & -4.0\%&+1.2\%&+0.3\%&-17.7\%& -18.2\%\\ 
B-RitualDance&-3.8\%  & -3.5\%  &-4.6\%  & -2.6\% &-1.1\%&-1.2\%&-11.7\%&-11.2\%\\ 
\midrule \textbf{Class B}&-3.1\% & -2.6\% &-4.5\% & -1.9\%&+0.3\%&-0.3\%&-15.3\%&-16.8\%\\ 
 \midrule C-BQMall&-5.6\% & -5.6\% &-4.7\% & -6.8\% &-2.1\%&-2.5\%&-14.3\%&-13.6\%\\ 
C-BasketballDrill&-3.9\% & -3.6\%  &-3.8\% & -2.8\% &-1.4\%&-1.6\%&-12.3\%&-11.7\%\\ 
C-PartyScene&-4.1\% & -4.3\%  &-5.9\% & -4.1\% &-0.4\%&-1.4\%&-16.1\%&-13.8\%\\ 
C-RaceHorses&-3.1\% & -2.1\%  &-3.4\% & +1.2\% &+0.2\%&-0.4\%&-11.9\%&-10.4\%\\ 
\midrule \textbf{Class C}&-4.2\%& -3.9\% &-4.5\% & -3.1\% &-0.9\%&-1.5\%&-13.7\%&-12.4\%\\ 
 \midrule D-BQSquare&-8.7\% & -9.6\% &-10.1\% & -11.6\%&-4.0\%&-4.5\%&-16.6\%& -19.5\%\\ 
D-BasketballPass&-6.1\% & -5.6\%  &-5.4\% & -4.0\% &-3.0\%&-2.8\%&-9.8\%&-8.1\%\\ 
D-BlowingBubbles&-3.7\% & -3.8\%  &-4.8\% & -3.8\% &-0.5\%&-1.3\%&-16.1\%&-14.5\%\\ 
D-RaceHorses&-4.8\% & -4.2\%  &-5.2\% & -1.0\% &-1.6\%&-1.9\%&-11.8\%&-10.5\%\\ 
\midrule \textbf{Class D}&-5.8\% & -5.8\% &-6.4\% & -5.1\% &-2.3\%&-2.6\%&-13.6\%&-13.2\%\\ 
 \midrule \midrule \multirow{2}{*}{\textbf{Overall}}&-4.0\% & -3.8\% &-4.9\% & -3.4\% &-0.6\%&-1.2\%&-13.5\%&-14.2\%\\ 
\cmidrule{2-9}&  \multicolumn{2}{c||}{BD-rate$=$-3.9\%}&  \multicolumn{2}{c||}{BD-rate$=$-4.2\%}&\multicolumn{2}{c||}{BD-rate$=$-0.9\%}&  \multicolumn{2}{c}{BD-rate$=$-13.9\%}\\
\bottomrule	
\end{tabular}
\label{tab:results_vtm}
\end{table*}

\begin{table*}[htbp]
\centering
\small
\caption{The compression performance of the proposed method benchmarked on the original AV1 1.0.0. Negative BD-rate values indicate coding gains.}
\begin{tabular}{l  || M{2cm}  ||M{2cm} ||M{2cm} ||M{2cm}}
\toprule
Method	& 	\multicolumn{2}{c||}{AV1-PP ($\ell 1$ loss)} & \multicolumn{2}{c}{AV1-PP (Perceptual loss)} \\
\midrule Metric&{PSNR} &{VMAF} & PSNR & VMAF \\ 
\midrule Class-Sequence & BD-rate & BD-rate & BD-rate & BD-rate \\
\midrule
A1-Campfire&-5.0\% &-7.3\% & -1.1\%  & -11.8\% \\ 
A1-FoodMarket4&-4.0\%  &-8.9\%  & -1.1\%  & -9.6\% \\ 
A1-Tango2&-4.6\%  &-8.2\% & -1.9\%  & -10.7\% \\ 
\midrule A2-CatRobot1&-5.9\%  &-5.9\%  & -1.9\%  & -15.0\% \\ 
A2-DaylightRoad2&-7.7\%    &-5.0\% & -3.1\%  & -17.2\% \\ 
A2-ParkRunning3&-1.9\%  &-2.2\% & +0.4\% &-11.3\%\\ 
\midrule \textbf{Class A}&-4.9\% &-6.3\% & -1.5\%& -12.6\% \\ 
 \midrule B-BQTerrace&-4.5\%  &+1.4\% & -1.3\%  & -10.2\% \\ 
B-BasketballDrive&-6.0\%   &-3.0\%  & -2.9\% &-7.1\% \\ 
B-Cactus&-3.8\%  &-3.3\%   &-0.7\%&-11.9\%\\ 
B-MarketPlace&-3.0\%  &-4.5\% & -0.5\% &-17.4\%\\ 
B-RitualDance&-4.7\%  &-5.0\%  & -2.3\% &-11.1\% \\ 
\midrule \textbf{Class B}&-4.4\% &-2.9\%& -1.5\%& -11.5\% \\ 
 \midrule C-BQMall&-6.3\% &-2.4\% & -2.9\% &-9.4\%\\ 
C-BasketballDrill&-6.8\%  &-2.4\% & -3.1\% &-9.9\%\\ 
C-PartyScene&-7.8\% & +2.3\% &-3.3\%&-9.1\%\\ 
C-RaceHorses&-4.1\% & -3.8\%  &-1.8\% & -8.4\% \\ 
\midrule \textbf{Class C}&-6.3\% &-1.6\% & -2.8\% &-9.2\% \\ 
 \midrule D-BQSquare&-16.1\% &+11.2\% & -8.4\% & -4.5\% \\ 
D-BasketballPass&-7.0\%  &-3.8\% & -3.9\% &-8.2\% \\ 
D-BlowingBubbles&-6.2\% &+2.0\% & -2.8\% &-9.7\% \\ 
D-RaceHorses&-5.6\% &-3.0\% & -3.0\% &-7.7\%\\ 
\midrule \textbf{Class D}&-8.7\% &+1.6\%& -4.5\% & -7.5\% \\ 
 \midrule \midrule \textbf{Overall}& -5.8\% &-2.7\% & -2.4\% & -10.5\% \\ 
\bottomrule	
\end{tabular}
\label{tab:results3_av1}
\end{table*}

The proposed post-processing approach has been utilised to enhance both VVC and AV1 compressed content. We have used all 19 test sequences from the JVET CTC standard dynamic range (SDR) testset. None of these sequences were included in the CNN training database.

VVC compressed content was generated using VTM 4.0.1 with the Random Access configuration. AV1 libaom 1.0.0 was used to produce AV1 content, with similar coding parameters to those for VVC. The employed configurations for both codecs are summarised in Table \ref{tab:cfg}. The tested QP values are 22, 27, 32, 37 and 42 for VVC, and 32, 43, 55 and 63 for AV1.

The final video quality was evaluated using two assessment methods, Peak Signal-to-Noise ratio (PSNR) and Video Multimethod Assessment Fusion (VMAF) \cite{w:VMAF}. PSNR is the most commonly used quality metric in the video compression community, while VMAF is a machine learning based metric, which combines multiple existing quality metrics and a video feature using a Support Vector Machine regression approach. Compared to PSNR, VMAF has been reported to provide more accurate prediction of perceptual quality. The performance of the proposed approach has been compared with two original codecs using the Bj{\o}ntegaard Delta rate measurements (BD-rate) \cite{r:Bjontegaard}. For VVC, the compression performance is evaluated for both low (22-37) and high QP (27-42) ranges.

In order to further benchmark the performance of the proposed algorithm, another two state-of-the-art CNN-based post-processing approaches, denoted as JVET-N0254 \cite{r:JVETN0254} and JVET-O0079 \cite{r:JVETO0079}, are also compared here in the context of VVC (for low QP range only). Results of both  are based on the RA configuration and were submitted to MPEG JVET meetings as VVC proposals.

\subsection{Compression performance}

\textbf{Table \ref{tab:results_vtm}} and \textbf{\ref{tab:results3_av1}} summarize the compression performance of the proposed method when it is applied to VVC and AV1 compressed content. For $\ell 1$ trained CNNs, we note that the average bit-rate savings according to PSNR are 3.9\% and 5.8\% against the original VVC and AV1 respectively. If the perceptual quality metric, VMAF, is used to assess video quality, the coding gains are 4.2\% over VVC and 2.7\% over AV1. When we use perceptual loss function trained models for post-processing, the coding gains appear much more significant based on the assessment of VMAF -- 13.9\% and 10.5\% over VVC and AV1 respectively. This is particularly evident for  test sequences, such as ParkRunning3, BQTerrace and BQSquare, which present a large number of sharp edges. We have also compared the proposed method with two JVET proposals, JVET-N0254 \cite{r:JVETN0254} and JVET-O0079 \cite{r:JVETO0079}, in \textbf{Table \ref{tab:vtmcompare}}, where the $\ell 1$ trained CNNs provides superior enhancement performance to these two works for all resolution classes according to PSNR.


\begin{table}[htbp]
\centering
\caption{Comparison between the proposed method and two existing CNN-based PP approaches for VVC (low QP range).}
\begin{tabular}{l || M{1cm} || M{1cm}|| M{1.15cm}}
\toprule
\multirow{2}{*}{Sequence (Class)} & {O0079} \cite{r:JVETO0079}&N0254 \cite{r:JVETN0254} &{Proposed Method} ($\ell 1$)\\
\cmidrule{2-4}
\centering
& BD-rate (PSNR) &   BD-rate (PSNR) &    BD-rate (PSNR)\\
 \midrule
 \centering
 \textbf{Class A (2160p)}& -1.3\%&-1.7\% 
&\textbf{-3.5\%}\\
\midrule \textbf{Class B (1080p)} &-1.5\%
& -1.1\% &\textbf{-3.1\%}\\
\midrule \textbf{Class C (480p)} &-3.3\%&
-1.4\% &\textbf{-4.2\%}\\
\midrule \textbf{Class D (240p)} &-5.0\%& -1.4\% &\textbf{-5.8\%}\\
  \midrule \textbf{Overall} &-2.6\%& -1.4\% &\textbf{-4.0\%}
\\\bottomrule
\end{tabular}
\label{tab:vtmcompare}
\end{table}

\subsection{Subjective comparison}

\begin{figure*}[ht]
\centering
\scriptsize
\centering
\begin{minipage}[b]{0.245\linewidth}
\centering
\centerline{\includegraphics[width=1\linewidth]{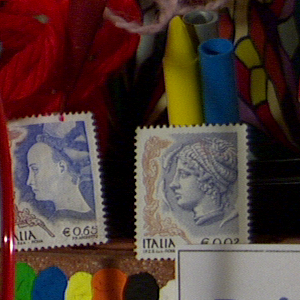}}
(a) Original \\ \ \ 
\end{minipage}
\begin{minipage}[b]{0.245\linewidth}
\centering
\centerline{\includegraphics[width=1\linewidth]{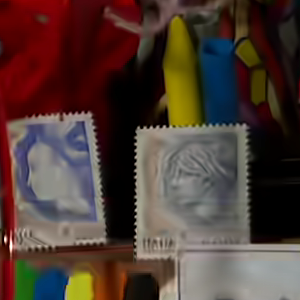}}
(b) VTM 4.0.1, QP=42 \\ (VMAF=62.5)
\end{minipage}
\begin{minipage}[b]{0.245\linewidth}
\centering
\centerline{\includegraphics[width=1\linewidth]{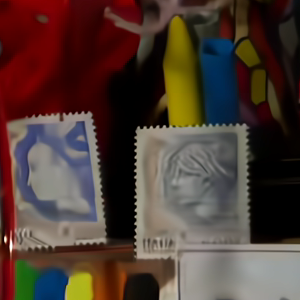}}
(c) VTM-PP: $\ell$1, QP=42 \\ (VMAF=64.0)
\end{minipage}
\begin{minipage}[b]{0.245\linewidth}
\centering
\centerline{\includegraphics[width=1\linewidth]{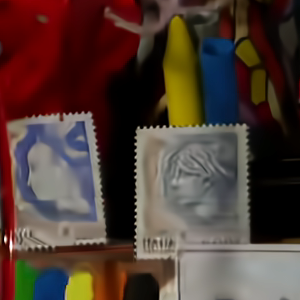}}
(d) VTM-PP: GAN, QP=42 \\ (VMAF=66.2)
\end{minipage}

\begin{minipage}[b]{0.245\linewidth}
\centering
\centerline{\includegraphics[width=1\linewidth]{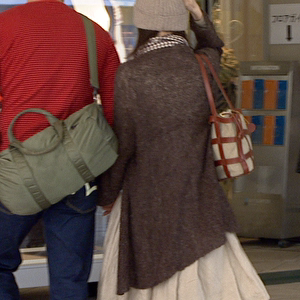}}
(e) Original \\ \ \ 
\end{minipage}
\begin{minipage}[b]{0.245\linewidth}
\centering
\centerline{\includegraphics[width=1\linewidth]{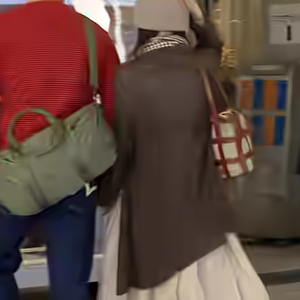}}
(f) VTM 4.0.1, QP=42 \\ (VMAF=70.4)
\end{minipage}
\begin{minipage}[b]{0.245\linewidth}
\centering
\centerline{\includegraphics[width=1\linewidth]{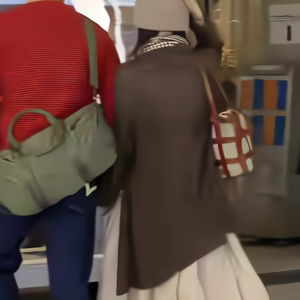}}
(g) VTM-PP: $\ell$1, QP=42 \\ (VMAF=72.2)
\end{minipage}
\begin{minipage}[b]{0.245\linewidth}
\centering
\centerline{\includegraphics[width=1\linewidth]{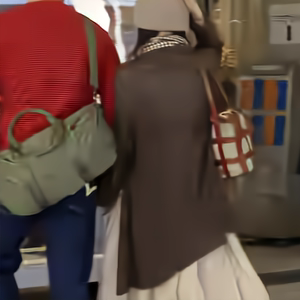}}
(h) VTM-PP: GAN, QP=42 \\ (VMAF=74.3)
\end{minipage}
\caption{Example blocks of the reconstructed frames for the anchor VTM 4.0.1 and the proposed approach. These are from the 91st and 320th frames of \textit{Cactus} and \textit{BQMall} sequences respectively. We can note that the results  produced by the perceptual loss function trained models exhibit more textural detail and higher contrast than those generated by $\ell$1 trained networks (the difference is more evident within the regions with the stamps in \textit{Cactus}, and the areas with the bags and the white skirt in \textit{BQMall}).}
\label{fig:perceptual1}
\end{figure*}

\begin{figure*}[ht]
\centering
\scriptsize
\centering
\begin{minipage}[b]{0.245\linewidth}
\centering
\centerline{\includegraphics[width=1\linewidth]{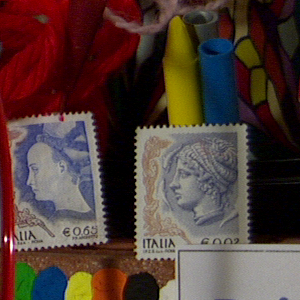}}
(a) Original \\ \ \ 
\end{minipage}
\begin{minipage}[b]{0.245\linewidth}
\centering
\centerline{\includegraphics[width=1\linewidth]{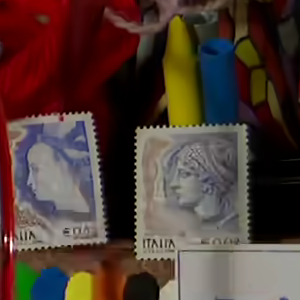}}
(b) AV1, QP=63 \\ (VMAF=75.2)
\end{minipage}
\begin{minipage}[b]{0.245\linewidth}
\centering
\centerline{\includegraphics[width=1\linewidth]{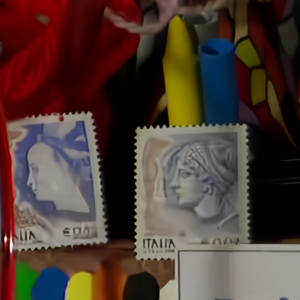}}
(c) AV1-PP: $\ell$1, QP=63 \\ (VMAF=76.4)
\end{minipage}
\begin{minipage}[b]{0.245\linewidth}
\centering
\centerline{\includegraphics[width=1\linewidth]{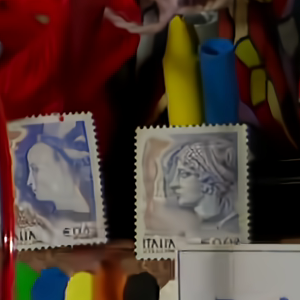}}
(d) AV1-PP: GAN, QP=63 \\ (VMAF=78.1)
\end{minipage}

\begin{minipage}[b]{0.245\linewidth}
\centering
\centerline{\includegraphics[width=1\linewidth]{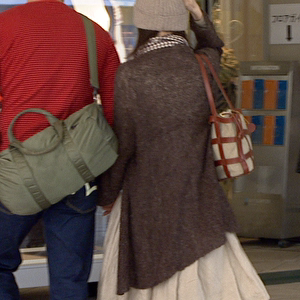}}
(e) Original \\ \ \ 
\end{minipage}
\begin{minipage}[b]{0.245\linewidth}
\centering
\centerline{\includegraphics[width=1\linewidth]{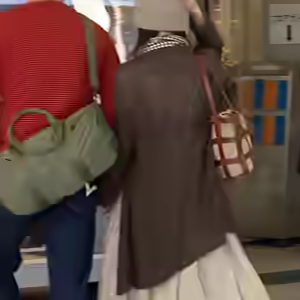}}
(f) AV1, QP=63 \\ (VMAF=83.7)
\end{minipage}
\begin{minipage}[b]{0.245\linewidth}
\centering
\centerline{\includegraphics[width=1\linewidth]{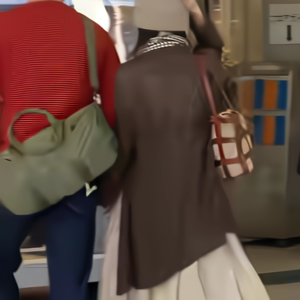}}
(g) AV1-PP: $\ell$1, QP=63 \\ (VMAF=84.9)
\end{minipage}
\begin{minipage}[b]{0.245\linewidth}
\centering
\centerline{\includegraphics[width=1\linewidth]{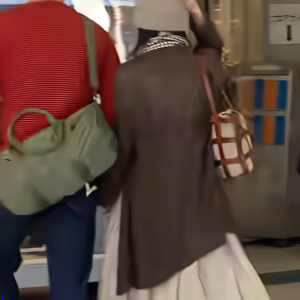}}
(h) AV1-PP: GAN, QP=63 \\ (VMAF=86.2)
\end{minipage}
\caption{Example blocks of the reconstructed frames for the anchor AV1 and the proposed approach. These are from the 91st and 320th frames of \textit{Cactus} and \textit{BQMall} sequences respectively. The results  produced by the perceptual loss function trained models exhibit more textural detail and higher contrast than those generated by $\ell$1 trained networks (the difference is more evident within the regions with the stamps in \textit{Cactus}, and the areas with the bags and the  white skirt in \textit{BQMall}).}
\label{fig:perceptual2}
\end{figure*}

\textbf{Figure \ref{fig:perceptual1}} and \textbf{\ref{fig:perceptual2}} provide subjective comparisons between  the reconstructed frames generated by the original VVC/AV1, $\ell 1$ trained CNNs and perceptual loss function trained models. We can observe that the reconstructed blocks of the proposed method (for both $\ell$1 and perceptually trained models) exhibit fewer noticeable blocking artefacts compared to the anchor codecs. In addition, the CNN models trained using perceptual loss functions produce results with slightly more textural detail and higher contrast than those generated by $\ell$1 trained networks.

\subsection{Complexity analysis}

The relative computational complexity of the proposed method, benchmarked on the original VVC and AV1 codecs, is presented in \textbf{Table \ref{tab:complexity}}. We have used a shared cluster computer at the University of Bristol, to execute all the computations. This computer contains multiple nodes with 2.4GHz Inter CPUs, 138GB RAM and NVIDIA P100 GPU devices. We note that the average decoding complexity is 56.3 and 70.0 times compared to the original VVC VTM 4.0.1 and AV1 respectively, due to the employment of CNN-based post-processing at the decoder. This is for a configuration with 16 residual blocks in the generator. This figure is slightly higher than those of two JVET proposals O0079 \cite{r:JVETO0079} and N0254 \cite{r:JVETN0254}, in which the decoder complexities (inc. the CNN-based PP module) are 45.6 and 35.2 times that of the original VVC decoder (based on the whole JVET dataset).

\begin{table}[htbp]
\centering
\caption{Relative Complexity of the proposed method benchmarked on original VVC and AV1 decoders.}
\begin{tabular}{l|| M{2cm}|| M{2cm}}

\toprule
\multirow{1}{*}{Hose Codec} & \multicolumn{1}{c||}{VVC} & \multicolumn{1}{c}{AV1}\\
\midrule
 \textbf{Class A (2160p)}&18.6$\times$&23.2$\times$\\
\midrule \textbf{Class B (1080p)} &36.8$\times$&38.3$\times$\\
\midrule \textbf{Class C (480p)} &76.5$\times$&83.3$\times$\\
\midrule \textbf{Class D (240p)} &116.9$\times$&166.7$\times$\\
  \midrule \textbf{Average} &56.3$\times$&70.0$\times$
\\\bottomrule
\end{tabular}
\label{tab:complexity}
\end{table}

\begin{figure}[htbp]
\centering
\centering
\begin{minipage}[b]{0.485\linewidth}
\centering
\centerline{\includegraphics[width=1.1\linewidth]{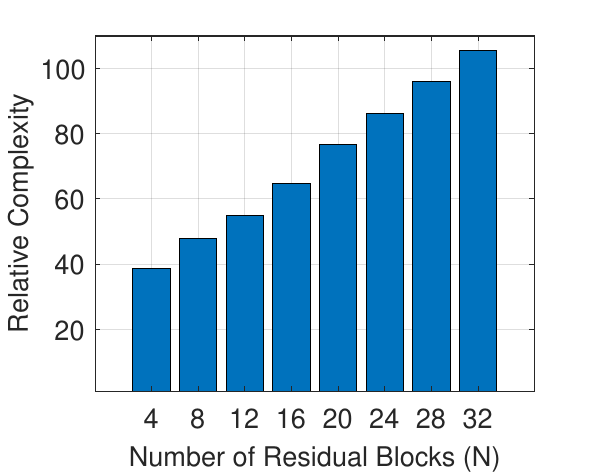}}
\end{minipage}
\begin{minipage}[b]{0.485\linewidth}
\centering
\centerline{\includegraphics[width=1.1\linewidth]{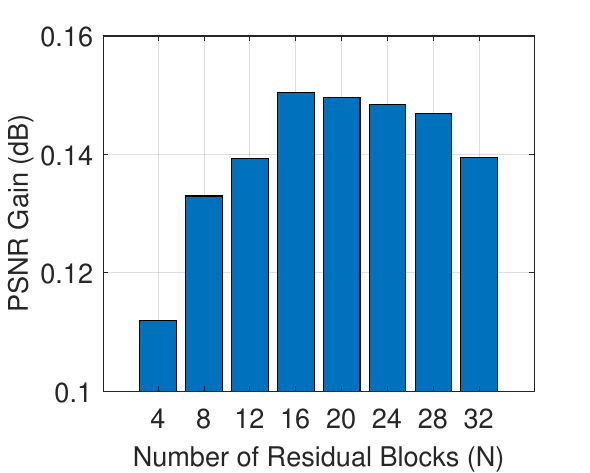}}
\end{minipage}
\caption{(Left)
Relative complexity for different number of residual blocks. (Right) PSNR gains for different number of residual blocks. }
\label{fig:bd}
\end{figure}

Moreover, we have further investigated the relationship between the number of residual blocks and compression performance. \textbf{Figure \ref{fig:bd}} shows the coding gains (in terms of PSNR) and algorithm relative complexity using CNN models with different numbers of residual blocks (N=4, 8, 12, 16, 20, 24, 28 and 32) to process VVC VTM QP 42 compressed content. Again the relative complexity here is benchmarked on the original VVC decoder. We observe that, when the number of residual blocks (N) increases from 4 to 16, the PSNR gain relative to the original VVC content increases in a linear fashion. However when the number of residual blocks exceeds 20, the overall coding gain starts to decrease. This provides evidence that the residual block number in the proposed work is an optimal selection.

\section{CONCLUSIONS}
\label{sec:conclusions}
 
In this paper, we presented a CNN-based post-processing approach, which achieves evident and consistent coding gains over standardized video codecs, VVC and AV1. The employed CNN model was trained using both $\ell 1$ and perceptually inspired methodologies. We would like to recommend future work focusing on computational complexity reduction and further improvement on the training methodology.
 
 \section{ACKNOWLEDGMENT}

We would like to acknowledge funding from UK EPSRC (EP/L016656/1 and EP/M000885/1) and the NVIDIA GPU Seeding Grants. We also appreciate the valuable advice provided by Dr. Debargha Mukherjee on AV1 coding configuration.

\bibliographystyle{IEEEtran}
\bibliography{IEEEabrv,MyRef}

\end{document}